# Quantum computation of lowest-energy Kramers states and magnetic g-factors of rare earth ions in crystals

K.M. Makushin, E.I. Baibekov

Kazan Federal University, 18 Kremlevskaya Str., Kazan 420008, Russia

We present the results of the quantum calculation of the ground state energies and magnetic g-factors of two rare earth (RE) ions: $Yb^{3+}$ in $Y_2Ti_2O_7$ crystal and $Er^{3+}$ in $YPO_4$ crystal. The Variational Quantum Eigensolver (VQE) algorithm has been performed on 5-qubit IBM superconducting quantum computer via IBM Quantum Experience cloud access. The Hamiltonian of the lowest spectroscopic multiplet of each RE ion, containing crystal field and Zeeman interaction, has been projected to the collective states of three ($Yb^{3+}$) and four ($Er^{3+}$) coupled transmon qubits. The lowest-energy states of RE ions have been found minimizing the mean energy in ~ 250 iterations of the algorithm: the first part performed on a quantum simulator, and the last 25 iterations - on the real quantum computing hardware. All the calculated ground-state energies and magnetic g-factors agree well with their exact values, while the estimated error of 2÷15% is mostly attributed to the decoherence associated with the two-qubit operations.

**I. Introduction**

During the past decade, the development of working physical realizations of multiqubit quantum computers has been quite rapid. In 2011, D-Wave One system was announced: a 128 qubit quantum annealing computer, which, however, was restricted to certain tasks and lacked the possibility to implement arbitrary quantum gates necessary to realize the "iconic" quantum algorithms [1]. In 2015, D-wave Systems announced the general availability of the D-Wave 2X, a 1000+ qubit quantum annealing computer. In 2016, IBM introduced their first superconducting quantum computer accessible by the vast scientific community through their web cloud [2]. In 2017, another company, Rigetti Computing, announced the public beta availability of a quantum cloud computing platform [3]. In 2019, Google claimed that their quantum 54-qubit processor achieved quantum supremacy [4].

However, quantum technology is still far enough from fault-tolerant computations, and most of algorithms we know are extremely sensitive to noise. Quantum computers with > 50 qubits may be able to perform tasks which surpass the capabilities of today's classical digital computers, but noise in quantum gates will limit the size of quantum circuits that can be executed reliably. In this sense, modern quantum devices are related to Noisy Intermediate-Scale Quantum (NISQ) technology [5]. Nevertheless, even now, it is sometimes difficult to simulate on a classical computer the algorithms that actually run on many-qubit hardware.

Among other applications, modern quantum computing platforms enable simulation of various coupled electronic quantum systems, including Ising and central spin models [6,7]. However, possible implementations obviously are not restricted to exchange-coupled spin aggregates. Another interesting problem tackled by the modern quantum computing devices is simulation of the lowest-energy state of simple atomic clusters like the hydrogen molecule [8]. A possible generalization would be a study of multi-electronic ion (a transition metal or rare earth ion) in crystal. Currently, we are unaware of any research on quantum computer simulations of spectroscopic properties of rare earth ions in crystals conducted so far. Each rare earth impurity ion has approximately 70 electrons coupled to each other, to their nucleus and to the nearby ions of the host crystal, which makes almost impossible to find an exact solution using any classical algorithm.

In this work, we deal with a more simple problem of finding the lowest-energy states of two Kramers ions ($Yb^{3+}$ and $Er^{3+}$), while taking into account only their lower-energy multiplet states. The electrostatic interaction between the electrons of the valence 4*f* shell, together with their spin-orbit interaction, produce a number of well-separated spectroscopic multiplets. The total angular momentum of 4f shell and its projection are considered as good quantum numbers. The electrostatic interaction produces some splitting between these levels within the multiplet, which can be modelled by introducing the so-called crystal field Hamiltonian [9]. The lowest-energy multiplets of $Yb^{3+}$ ($^2F_{7/2}$) and $Er^{3+}$ ($^4I_{15/2}$) contain 8 and 16 states, respectively, can be effectively simulated by using the states of 3 or 4 coupled qubits, respectively.

**II. Methods**

*II.1. Crystal field and spectroscopic data*

During the last two decades, the pyrochlore crystals with various impurity RE ions, being common physical realizations of a geometrically frustrated system, attracted attention of many researchers [10]. Here we consider $Y_2Ti_2O_7$ crystal doped with $Yb^{3+}$ ions containing 13 electrons on the valence 4*f* shell. Their electrostatic interaction with the surrounding $Y^{3+}$, $Ti^{4+}$ and $O^{2-}$ ions of the host crystal projected on the lowest-energy multiplet $^2F_{7/2}$ (orbital momentum $L = 3$, spin $S = 1/2$, total angular momentum $J = 7/2$) produces the effective crystal field operator, which, in the case of trigonal symmetry of the $Yb^{3+}$ site, is defined as:

$$H_{\mathrm{CF}}^{(\mathrm{Yb})} = \alpha_2 B_2^0 O_2^0 + \alpha_4 \left( B_4^0 O_4^0 + B_4^3 O_4^3 \right) + \alpha_6 \left( B_6^0 O_6^0 + B_6^3 O_6^3 + B_6^6 O_6^6 \right). \qquad (1)$$

Above, $O_p^k$ are Stevens operators (linear combinations of spherical tensor operators [9,11]), the parameters $\alpha_p$ define the reduced matrix elements [9]. The crystal field parameters $B_p^k$ determined previously [11] are presented in Table 1.

The second system that we consider in this work is yttrium orthophosphate crystal $YPO_4$ activated with $Er^{3+}$ ions. This system is a promising telecom-wavelength material for applications in quantum electronics and quantum information processing [12]. $Er^{3+}$ ions substitute for $Y^{3+}$ ions on the sites of $D_{2d}$ symmetry. The crystal-field interaction projected onto the subspace of $^4I_{15/2}$ multiplet ($L = 6$, $S = 3/2$, momentum $J = 15/2$) is expressed as:

$$H_{\mathrm{CF}}^{(\mathrm{Er})} = \alpha_2 B_2^0 O_2^0 + \alpha_4 \left( B_4^0 O_4^0 + B_4^4 O_4^3 \right) + \alpha_6 \left( B_6^0 O_6^0 + B_6^4 O_6^4 + B_6^6 O_6^6 \right). \qquad (2)$$

The parameters $B_p^k$ determined previously [12] are presented in Table 1.

Table I. Crystal field parameters $B_p^k$ (cm$^{-1}$) and the reduced matrix elements $\alpha_p$ for Yb$^{3+}$ in Y$_2$Ti$_2$O$_7$ crystal and Er$^{3+}$ in YPO$_4$ crystal.

| p | k | Yb$^{3+}$ | | Er$^{3+}$ | |
|---|---|---|---|---|---|
| | | $\alpha_p$ | $B_p^k$ [11] | $\alpha_p$ | $B_p^k$ [12] |
| 2 | 0 | $\dfrac{2}{63}$ | 264.8 | $\dfrac{4}{1575}$ | 112.9 |
| 4 | 0 | $-\dfrac{2}{1155}$ | 270.8 | $\dfrac{2}{45045}$ | 10.3 |
| | 3 | | –2155.2 | | |
| | 4 | | | | 776 |
| 6 | 0 | $\dfrac{4}{27027}$ | 44.9 | $\dfrac{8}{3864861}$ | –43.0 |
| | 3 | | 636.6 | | |
| | 4 | | | | 56.1 |
| | 6 | | 683.2 | | 112.9 |

Interaction with the static magnetic field projected onto the ground multiplet $^{2S+1}L_J$ results in the following Zeeman interaction term:

$$H_Z = g_J \mu_B \mathbf{JB}, \qquad (3)$$

where $\mu_B$ is Bohr magneton, **B** is the magnetic field vector, and $g_J$ is Landé g-factor (below, $g_e = 2.0023$ is electronic g-factor) [9]

$$g_J = (g_e - 1) \cdot \frac{J(J+1) + S(S+1) - L(L+1)}{2J(J+1)} + 1. \qquad (4)$$

The matrices of the operators (1)-(3) in the basis of $|J, M_J\rangle$ states have been computed on a classical computer, and then expanded into series of either three- or four-fold tensor products of Pauli matrices, see Section II.3.

*II.2. Computing hardware and software*

We use cloud access to IBM devices for our experiments trough the Qiskit framework [13]. IBM processors utilize transmon qubit technology [14,15]. Transmon qubit is a form of superconducting charge qubit with additional capacity resistant to charge noise. The device comprises of fixed-frequency Josephson-junction-based transmon qubits, with individual superconducting coplanar waveguide (CPW) resonators for qubit control and readout, and another pair of CPW resonators providing the qubit connectivity. This fixed-frequency architecture is favorable for obtaining long coherence times, and the qubit control and readout is achieved using only microwave pulses. Each quantum chip is calibrated on a daily basis, thus minimizing its single-qubit and CNOT error rates. Since our problems do not involve more than 4 qubits, we mainly use a 5-qubit device depicted in Fig. 1.

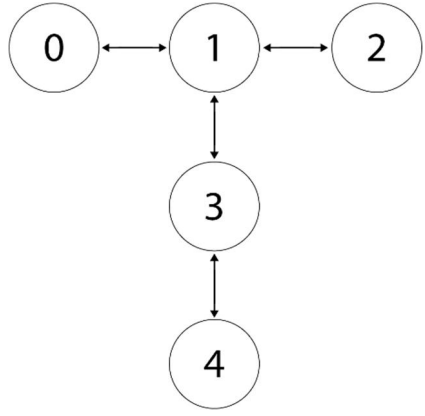

Figure 1. Layout of a 5-qubit IBM quantum computer. The arrows indicate the coupling arranged between the adjacent qubits.

A quantum algorithm for the ground-state problem proposed in works [16,17] relies on preparation of the initial state that has a large overlap with the ground state and solves the problem using the quantum phase estimation algorithm (PEA). This approach requires sufficient circuit depths and long coherence times, and such requirements cannot be fully handled by contemporary hardware. In our work, we use the VQE algorithm introduced in 2014 by Peruzzo and others [18,19], which has less strict hardware requirements and is suitable for eigenvalue calculations using NISQ devices. The low-depth circuits that VQE utilizes can help to avoid decoherence errors during computation. The algorithm is based on Ritz variational principle. If we parametrize by $\boldsymbol{\theta}$ an eigenfunction $|\psi(\boldsymbol{\theta})\rangle$ of the Hamiltonian $H$, and minimize left side of inequality (5), we will be able to approach the ground state energy $E_0$ of $H$:

$$\frac{\langle\psi(\boldsymbol{\theta})|H|\psi(\boldsymbol{\theta})\rangle}{\langle\psi(\boldsymbol{\theta})|\psi(\boldsymbol{\theta})\rangle} \geq E_0. \tag{5}$$

Since VQE is a semi-classical algorithm, it consists of two parts: (i) the quantum-computer part that evaluates the expectation value (5) of the target Hamiltonian, and (ii) the classical-computer part that seeks minimum of the expectation value with respect to $\boldsymbol{\theta}$ using a classical optimization algorithm.

In order to parametrize the eigenstate, we have to adopt one of the strategies which are used to build so-called ansatz circuits to prepare initial or trial state. First one relies on the knowledge of properties and symmetries of the system's Hamiltonian. Such ansatzes are widely used in quantum chemistry problems, e.g. the Unitary Coupled Cluster Ansatz (UCCA) in recent studies [20-22]. Second strategy is called Hardware-Efficient Ansatz: this one utilizes quantum gates that are naturally available on the quantum hardware [23]. We have chosen second strategy for our work, because it requires less circuit depth, as we should exclude as many error sources as possible due to limited coherence times of contemporary IBM quantum devices.

Let us briefly summarize the basic principles of VQE approach [18,19]. For the initial state $|\varphi\rangle$, it would be convenient to choose one of the states of the computational basis $\{|00\ldots0\rangle, |00\ldots1\rangle, \ldots |11\ldots1\rangle,\}$ $\{|00\ldots0\rangle, |00\ldots1\rangle, \ldots |11\ldots1\rangle\}$. Our ansatz acts on this state as unitary transformation $U(\boldsymbol{\theta})$:

$$|\psi(\boldsymbol{\theta})\rangle = U(\boldsymbol{\theta})|\varphi\rangle, \qquad (6)$$

where $\boldsymbol{\theta} = \{\theta_1, \theta_2, \ldots, \theta_m,\}$ is a set of $m$ parameters. A generalized unitary transformation performed on a single qubit modeling the states of a spin-1/2 particle could be written as a combination of three spin rotations around $z$ and $y$ axes in the form [24]:

$$U^{q,i}(\boldsymbol{\theta}) = R_Z\left(\theta_1^{q,i}\right) R_Y\left(\theta_1^{q,i}\right) R_Z\left(\theta_3^{q,i}\right), \qquad (7)$$

where $q$ identifies the qubit and $i$ identifies the algorithm depth layer of the circuit. For $N$ qubits, we can write:

$$U(\boldsymbol{\theta}) = \prod_{q=1}^{N} U^{q,d}(\boldsymbol{\theta}) U_{\text{ENT}} \prod_{q=1}^{N} U^{q,d-1}(\boldsymbol{\theta}) U_{\text{ENT}} \cdots \prod_{q=1}^{N} U^{q,0}(\boldsymbol{\theta}), \qquad (8)$$

where $U_{\text{ENT}}$ represents an entangling layer of two-qubit gates. This gives the algorithm $N(3d+3)$ parameters to be optimized, where $d$ is the depth of the algorithm. After each trial state is prepared, we estimate the associated mean energy (5) by measuring the expectation values of the individual Pauli tensorial products $\langle P \rangle = \langle A^1 \otimes B^2 \otimes \cdots \otimes C^N \rangle$, where each $A, B, \ldots, C$ is either one of three Pauli operators $X^q, Y^q, Z^q$ of the qubit $q$, or identity operator $I^q$. The Hamiltonian $H$ is presented as a sum of such products, see section II.3. Since we work in the basis of the eigenstates of $Z^q$ operators, then $Z^q$ values can be obtained by directly measuring the final state of a qubit $q$, and then averaged over a large number of runs (8192 in our case) to produce $\langle \ldots Z^q \ldots \rangle$. In order to make measurements for another two Pauli operators $X^q, Y^q$, we apply additional $\pi/2$ rotations prior to measurement of $Z^q$.

After preparation of the trial state is finished and all the measurements are done, the algorithm continues with its classical part. It suggests the summation of all average values $\langle P \rangle$ comprising the expectation value of $H$, and a search of a better set of parameters that would lower $\langle H \rangle$. After that, the whole cycle repeats again until the convergence conditions are fulfilled.

*II.3. Quantum computation of the lowest energy state of a RE ion*

In order to conform with VQE approach, we must map the problem's Hamiltonian $H$ to the $N$-qubit Hamiltonian containing Pauli tensorial products. We use Hilbert-Schmidt inner product decomposition:

$$H = \frac{1}{N^2} \sum_{i,j,\ldots,k} h_{i,j,\ldots,k} \cdot X_i^1 \otimes X_j^2 \otimes \cdots \otimes X_k^N, \quad h_{i,j,\ldots,k} = \frac{1}{N} \text{Tr}\left(X_i^1 \otimes X_j^2 \otimes \cdots \otimes X_k^N \cdot H\right). \qquad (9)$$

Above, $i, j, \ldots, k$ refer to $\{X, Y, Z, I\}$ operators. In the case of $Yb^{3+}$ ion's $^2F_{7/2}$ state, we map its Hamiltonian (1) to 8 states of three qubits, and obtain (omitting the upper indices for simplicity)

$$H_{CF}^{(Yb)} = h_1 IZZ + h_2 XXI + h_3 XZX + h_4 YYI + h_5 YZY + h_6 ZIZ + h_7 ZXX + h_8 ZYY + h_9 ZZI. \quad (10)$$

Next, we can utilize the fact that the crystal field Hamiltonian contains only real terms. In this case, one can simplify the one-qubit rotation (7) to the form that contains only one parameter $U^{q,i} = R_Y(\theta^{q,i})$ [25]. With this simplification, our algorithm now has only $N(d+1)$ parameters to be optimized.

For the classical optimization part, we choose Simultaneous Perturbation Stochastic Approximation (SPSA) algorithm, since it is robust to stochastic fluctuations and requires only two cost-function evaluations [23], irrespective of the dimensionality of the parameter space. Because we did not have continuous access to IBM quantum devices, to avoid time consuming, the first part of the optimization procedure has been performed on the simulator of a quantum computer. The energy was estimated on the real hardware during the last 25 iterations of optimization, and then averaged over several iterations to obtain better results.

It is possible to reduce the number of terms in the Hamiltonian using simple qubit-wise commutativity (QWC) [26] for Pauli operators. Two Pauli strings QWC-commute if, at each index, the corresponding two Pauli operators commute. For instance, $\{XX, XI, IX, II\}$ is a QWC partition, so all these Pauli strings can be measured simultaneously, and the results then re-calculated straightforwardly to obtain the corresponding expectation values. Since each additional measurement is likely to increase the error, the use of QWC is advantageous. To this extent, we can rewrite the Hamiltonian (10) in the form

$$H_{CF}^{(Yb)} = (h_1 + h_6 + h_9) IZZ + h_2 XXI + h_3 XZX + h_4 YYI + h_5 YZY + h_7 ZXX + h_8 ZYY. \quad (11)$$

For $Er^{3+}$ ion, the Hamiltonian (2) maps to the Hilbert space of four qubits:

$$\begin{aligned} H_{CF}^{(Er)} = &h_1 IIZZ + h_2 IXII + h_3 IXZZ + h_4 IZIZ + h_5 IZZI + h_6 XXII + h_7 XXZZ + h_8 YYII + \\ &+ h_9 YYZZ + h_{10} ZIIZ + h_{11} ZIZI + h_{12} ZXIZ + h_{13} ZXZI + h_{14} ZZII + h_{15} ZZZZ. \end{aligned} \quad (12)$$

Again, using QWC approach, we reduce the number of independently measured terms to 4. The Zeeman interaction (3) has been decomposed into the Pauli strings in the same manner. The magnetic g-factors of the lowest Kramers doublet of each multiplet have been calculated by minimizing the energy of the lowest state of the ion subjected to the external magnetic field **B** directed along ($g_\parallel$) or perpendicular to ($g_\perp$) the crystal's symmetry axis. In either case,

$$g = \frac{2\langle \psi(\boldsymbol{\theta}^*)|H_Z|\psi(\boldsymbol{\theta}^*)\rangle}{\mu_B B}, \quad (13)$$

where $\boldsymbol{\theta}^*$ represents the parameter set minimizing the expectation value of total Hamiltonian $H_{CF} + H_Z$. We have also calculated the standard error

$$\alpha = \frac{\sigma}{\sqrt{n}}, \quad (14)$$

where $\sigma$ is a standard deviation of the result, $n$ is the number of measurement repetitions. The standard deviation in measurement of one of the Pauli strings $P_i$ in the Hamiltonian is given by

$$\sigma_{P_i} = \sqrt{\langle \psi | P_i^2 | \psi \rangle - \langle \psi | P_i | \psi \rangle^2} = \sqrt{1 - \langle \psi | P_i | \psi \rangle^2} \leq 1. \qquad (15)$$

The standard error in the energy measurement is then upper bounded by [23]

$$\alpha_E = \sqrt{\sum_i \alpha_{p_i}^2} = \sqrt{\sum_i \frac{h_i^2 \sigma_{p_i}^2}{n_i}} \leq \sqrt{\sum_i \frac{h_i^2}{n_i}}. \qquad (16)$$

This quantity is shown in the inset of Fig. 2.

**III. Results and discussion**

Some of the results obtained for the YPO$_4$:Er$^{3+}$ crystal are shown on Fig.2. The SPSA minimization was performed with Qiskit qasm_simulator, and the ground-state energy has been calculated on the real hardware during the last 25 steps of optimization with 5-qubit quantum chip. The exact eigenvalues were calculated with standard diagonalization procedure. As shown in Fig. 2, the difference between the final energy estimate and the "exact" ground state energy is less than 10%. The calculated g-factor error values are within the range of 2÷15%, see Table II. The estimated error is attributed to the decoherence associated with the two-qubit operations.

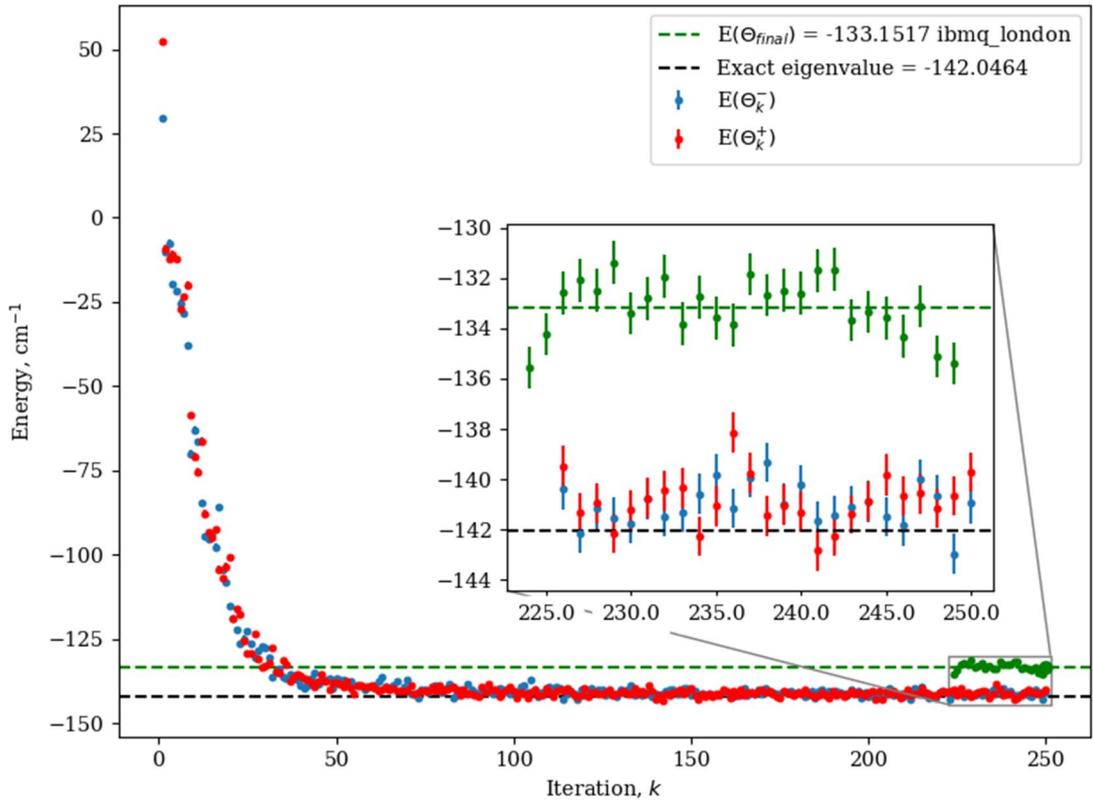

Figure 2. Results of the minimization procedure in the case of $Er^{3+}$ crystal field Hamiltonian. The first part (red and blue dots) correspond to the quantum simulator results obtained for the two close sets of parameters determining the next-iteration set in SPSA. The last part of the curve (green dots) corresponds to the quantum hardware part of the calculation.

Table II. The g-factors of the lowest-energy Kramers doublet.

|  | $Yb^{3+}$ in $Y_2Ti_2O_7$ | | | $Er^{3+}$ in $YPO_4$ | | |
|---|---|---|---|---|---|---|
|  | Experiment [11] | Classical diagonalization | Quantum calculation | Experiment [12] | Classical diagonalization | Quantum calculation |
| $g_\parallel$ | 1.787 | 1.864 | 1.566±0.058 | 6.42 | 6.78 | 6.463± 0.027 |
| $g_\perp$ | 4.216 | 4.181 | 4.153±0.079 | 4.80 | 4.71 | 5.097±0.063 |

**Acknowledgment**

The work was performed with the support from the subsidy allocated to Kazan Federal University for the state assignment in the sphere of scientific activities No. FZSM-2020-0050.

Authors also acknowledge the support of the Russian Science Foundation, Project No. 17-72-20053.